\documentclass[pra,superscriptaddress,showpacs,nobalancelastpage]{revtex4}

\usepackage{graphicx}
\begin{document}

\title{Decoherence control in microwave cavities}

\author{Stefano Zippilli}
\author{David Vitali} 
\author{Paolo Tombesi} 
\affiliation{Dipartimento di Fisica and Unit\`a INFM, 
Universit\`a di Camerino, \\ 
via Madonna delle Carceri I-62032 Camerino, Italy}
\author{Jean-Michel Raimond}
\affiliation{Laboratoire Kastler Brossel, D\'epartement de Physique
de l'Ecole Normale Sup\'erieure, \\ 
24 rue Lhomond, F-75231 Paris Cedex 05, France
}

\date{\today}

\begin{abstract}
We present a scheme able to protect the quantum states 
of a cavity mode against the decohering effects of photon 
loss. The scheme preserves quantum states 
with a definite parity, and improves
previous proposals for decoherence 
control in cavities. It is implemented by sending single atoms, one by one,
through the cavity.
The atomic state gets 
first correlated to the photon number parity. 
The wrong parity results in an atom in the 
upper state. The atom in this state is then used to inject 
a photon in the mode via adiabatic transfer, correcting the field parity.
By solving numerically the exact master equation of the 
system, we show that the protection of simple quantum states could be 
experimentally demonstrated using presently available experimental 
apparatus.
\end{abstract}


\maketitle

\section{Introduction}

In recent years, considerable effort has been devoted 
to designing strategies 
able to counteract the undesired effects of the coupling with an
external environment. Notable examples are quantum error correction codes
\cite{qecc} and error avoiding codes \cite{eac}, both based on 
encoding the state to be protected into 
carefully selected subspaces of the joint Hilbert space of the system 
and a number of ancillary systems. 
The main limitation for the efficient implementation of these 
encoding strategies for combating decoherence is the 
large amount of extra resources required \cite{steane}. Correcting all 
possible one-qubit errors requires at least five qubits 
\cite{zurprl}. 
This number rapidly increases if fault tolerant error correction 
is also considered. 
For this reason, other alternative 
approaches which do not require ancillary resources have been 
pursued and developed in parallel with encoding strategies.
These decoherence control schemes may be divided into two main categories: 
open-loop 
\cite{viola,ban,noi,agarwal,berman,termico}, 
and closed-loop (or quantum feedback) 
strategies \cite{closed,mabuchi,protect,jmr}. 

In open loop techniques (also called dynamical decoupling schemes),
the system is subject to an external, 
suitably tailored, time-dependent driving. The external control 
Hamiltonian is chosen on the basis of a limited, {\em a priori}, 
knowledge of the system-environment dynamics, 
in order to realize an effective 
dynamical decoupling of the system from the environment. 
The main idea 
behind these open loop schemes 
originates in refocusing techniques 
in NMR spectroscopy \cite{waugh}, but
they have been recently 
transposed in many different contexts, such as the 
inhibition of the decay of an unstable atomic state 
\cite{agarwal}, the suppression of magnetic state decoherence 
\cite{berman} and the reduction of heating effects in linear ion 
traps \cite{termico}. 
The main drawback of open loop decoupling procedures is that the 
timing constraints are particularly stringent. In fact, the decoupling 
interaction has to be turned on and off at extremely short time 
scales, even faster than typical environmental timescale 
\cite{viola,noi,termico}. 
An approach to quantum state protection related to decoupling schemes 
is represented by reservoir engineering schemes \cite{poyatos,lutk,david},
in which an external 
driving is used to create an effective reservoir for the system. 
In such a way, the state to be protected becomes a stationary state
of the modified dynamics. Examples have 
been proposed for the center-of-mass motion of trapped ions 
\cite{poyatos,david} and for atomic internal states \cite{lutk}.

It is interesting to notice that quantum error correction codes, error 
avoiding codes and decoupling schemes can be described in a unified 
framework based on the representations of the algebra of errors (the 
algebra generated by the set of system operators describing the 
effects of the environment) \cite{zana2,viola4}. 
In the error-algebra framework, quantum information is protected using 
symmetry. In the case of decoherence-free subspaces the symmetry 
naturally exists in the interaction with the environment. In the case 
of decoupling techniques, symmetry is induced by the added driving 
Hamiltonian. Finally, in quantum error correction codes, symmetry exists 
implicitly within the larger Hilbert space of the system and
ancillae \cite{zana2}.

Closed loop techniques represented the first attempt to control 
decoherence \cite{closed,mabuchi}. In this case, the system
to be protected is subject to appropriate measurements, and the 
classical information obtained from this measurement is used for 
real-time correction of the system dynamics. This technique shares 
therefore some similarities with quantum error correction, which also 
checks which error has taken place and eventually corrects it. 
However, the main limiting aspect of feedback schemes is the need of 
a measurement, which is always inevitably 
prone to errors and finite detection efficiency. For this reason, recent 
attempts have tried to improve these closed-loop schemes by avoiding 
the explicit measurement step, resorting to what may be called 
``fully quantum feedback'' schemes, in which 
sensors, controller, and actuators are themselves quantum systems 
and interact coherently with the system to be controlled \cite{lloyd}.
In this case, the entire feedback loop is coherent and is not 
limited by measurement inefficiencies. In fact, some of us have 
already proposed a scheme of this kind \cite{jmr}. It improved an 
existing closed-loop scheme for decoherence control \cite{protect},
by replacing the measurement step (an atomic 
detection) with the coherent interaction with a 
quantum controller, represented by a high-Q cavity.
The quantum coherent interaction allows
an automatic correction of the dynamics, without needing
an explicit measurement, similar to what happens in
quantum error correction codes.

In this paper we proceed further along this direction by introducing a 
significant simplification of the ``automatic'' scheme of \cite{jmr}. 
As in \cite{jmr}, the present 
scheme is designed to protect an arbitrary quantum state of a microwave cavity mode
with a given parity against the decohering 
effects of photon loss. 
However, in the present proposal, the whole feedback loop is realized by a 
single atom crossing the cavity.  
It ``measures'' the parity of the 
field in the first part of its interaction with the cavity.
The atom then performs, when needed, the 
state correction in the second part of the interaction time. For this reason, 
this scheme is another example of a fully quantum feedback 
loop \cite{lloyd}, which employs very limited 
resources, namely a single atom playing
the roles of sensor, controller and actuator.
In the scheme of Ref.~\cite{jmr} instead, 
a first atom is the sensor, a second 
high-$Q$ cavity is the controller. A second atom is used as the actuator.
The proposed strategy shares also some analogies
with quantum error correction codes. 
The error to be corrected is the loss of one photon, 
which drives the state out of a parity eigenspace.
With this respect, the two-level atom implementing the scheme plays 
the role of the error syndrome, because its state denotes the 
eventual presence of an error, that is, of a wrong parity. 

The simplifications brought by the present scheme are relevant
since they make the present proposal much easier to implement than 
that of \cite{jmr}, because it does not need a second high-$Q$ cavity and 
the use of a second atomic source. 
An easy implementation is an important asset. In spite of 
very many theoretical proposals for decoherence control, 
there has been only very few experimental demonstrations. 
A simple example of decoherence-free subspace immune from magnetic 
field noise has been demonstrated with two trapped ions in 
\cite{kielp}, while error correction 
codes for single qubit errors 
has been demonstrated only in NMR quantum information processors 
\cite{nmrqecc}. Outside the usual application in NMR refocusing 
techniques, dynamical decoupling schemes have been
implemented only in Ref.~\cite{bergl},
where a proof-of-principle demonstration for
a photon polarization qubit with artificially added decoherence has 
been realized (see also Ref.~\cite{fortu} for a recent 
demonstration of {\em encoded} decoupling schemes in NMR systems). 
A simple demonstration of fully quantum feedback has been given 
in the case of a three-nuclear spin system in 
\cite{lloyd2}, but no closed-loop decoherence control 
scheme has been demonstrated yet. 
The eventual implementation of the present proposal 
is important also because it would represent the first demonstration 
of the control of an intrinsic and unavoidable decoherence source,
photon loss,
instead of the control of an added noise.

The paper is organized as follows: In Section II, 
the cavity QED model under study is described, and the quantum state 
protection scheme is presented in detail. In Section III, the 
performance of the proposed scheme is studied by solving 
numerically the dynamical evolution of the system. Different examples 
of initial quantum states of the radiation mode to protect will be 
considered. Section IV is for concluding remarks.

\section{The state protection scheme}

The general purpose of decoherence control schemes is to protect 
a given subspace of a system Hilbert space, 
and quantum coherent evolutions within it. 
The conditions under which these noiseless subsystems 
exist, and universal quantum computation within them is possible have 
been already illustrated in the recent literature, especially in the 
case of qubits \cite{viola,zana2,viola4,vkl,lidar}. However, the 
experimental realization of these general schemes 
is difficult in many physical situations.
This is particularly true in infinite dimensional systems as 
radiation modes, which are of fundamental importance for any 
quantum communication scheme and for which only few specific 
(and difficult to implement) quantum error correction
schemes have been proposed 
\cite{cvqecc}. 
Here, we shall focus on the case of a radiation mode confined 
in a cavity. The first examples of quantum gates and 
quantum state manipulations have been demonstrated
in this context \cite{disp,turchette,cat,noncl,varcoe,mani}. 
Moreover, cavity modes could represent
the nodes of a quantum network of multiple 
atom-cavity systems linked by optical interconnects \cite{pell}.

In electromagnetic cavities, decoherence is mainly of dissipative 
origin and it is associated with the photon losses due to diffraction
and to the transmission and absorption of the mirrors.
In the general case where the reservoir of the continuum of 
electromagnetic modes is at thermal equilibrium at temperature $T$, 
the dynamics is well described by the master equation (in the frame 
rotating at the mode frequency $\omega$) \cite{milwal}
\begin{equation}
\dot{\rho }= {\cal L}(a)\rho \equiv \frac{\gamma}{2}(N+1) 
\left(2a \rho a^{\dagger} -a^{\dagger}a \rho - \rho 
a^{\dagger}a\right) +
\frac{\gamma}{2}N 
\left(2a ^{\dagger} \rho a -aa ^{\dagger} \rho - \rho 
aa^{\dagger}\right) \;,
\label{meq}
\end{equation}
where $\rho$ is the field density matrix and $a$ the annihilation operator of the 
cavity mode, $\gamma$ the cavity decay rate and $N=\left[\exp(\hbar 
\omega/kT)-1\right]^{-1}$ is the equilibrium thermal photon number. 
The cavity mode 
is affected by two kinds of errors, photon loss (with rate $\gamma 
(N+1)$) and thermal photon creation (with rate $\gamma N$). However,
in many cases, one has $N \ll 1$. Photon loss is then by far 
the predominant source of decoherence. 

In \cite{protect}, a 
closed-loop scheme for protecting a generic state of a cavity mode has been
proposed, based on the simple idea of giving back the photon as soon 
as it is lost. In the case where the sensor is represented by
a single-photon photodetector with quantum efficiency $\eta$ continuously
monitoring the cavity, the dynamics in the presence of feedback is
described by the master equation (in the case $N\simeq 0$) \cite{protect}
\begin{equation}
\dot{\rho }= (1-\eta)\frac{\gamma}{2} 
\left(2a \rho a^{\dagger} -a^{\dagger}a \rho - \rho 
a^{\dagger}a\right) -
\eta \frac{\gamma}{2} 
\left[\sqrt{a^{\dagger}a},\left[\sqrt{a^{\dagger}a},\rho\right]\right] \;.
\label{meqfeed}
\end{equation}
In the case of perfect detection ($\eta =1$)
cavity damping is therefore replaced by an unconventional phase-diffusion
process. In the ideal case, the only well-preserved states are the Fock states.
However, since the phase diffusion
process is very slow, the resulting quantum state protection is still
significant for other states \cite{protect}.

In the case of microwave cavities, there are no efficient single photon
detectors besides atoms crossing the cavity mode. The photon counting
can be replaced in this case by a field parity 
measurement\cite{protect,engl,brune96,lutt,wign}.
This measurement can be efficiently performed by using a 
dispersive atom-field interaction revealed by
a Ramsey interferometry set-up \cite{Ramsey}.
If these parity measurements are repeated at short time intervals,
so that multiple photon losses between them are negligible, they
reveal unambiguously the photon losses and
replace a single-photon photodetector. This is precisely the 
stroboscopic measurement scheme proposed in \cite{protect} 
for the cavity QED microwave experiments described in details in \cite{rmp}.
The price to pay when using parity measurements instead of photon counting 
is that only states with a given
parity can be protected.

Ref.~\cite{jmr} improved this closed-loop decoherence control scheme 
by transforming it into one of the first examples of ``fully quantum 
feedback loop''. 
In \cite{protect}, the feedback loop involved a
first atom probing the parity of the cavity mode. 
The final state of the first atom, correlated to the field parity, was 
measured by a state-selective atomic detector. 
Depending upon the result of this measurement, 
a dedicated electronics could send a second
atom through the cavity. This atom would emit a photon in the mode, 
correcting the effect of photon
loss and restoring the initial field parity.
Therefore, in this case, both the sensor (first atom +
detector) and the controller (the electronics) are essentially classical, 
while only the actuator (the second atom) is a quantum system.
In \cite{jmr} the detector and the controlling electronics are replaced by
a second high-$Q$ microwave cavity, resonantly interacting with the two atoms.
This cavity becomes the controller and the feedback loop has become completely
quantum. Here, we propose a further improvement of this protection
scheme, making its experimental implementation easier.
The present scheme is based again on the measurement of the cavity mode parity,
but it involves only one atom, which, in passing through the cavity,
first measures and then corrects the state of the mode when needed.
The simplification of the design is evident, with a single
atom realizing the whole loop, by playing all the roles of sensor, controller
and actuator.

\subsection{The physical system and the protection scheme in detail}

The microwave cavity QED set-up which we have specifically considered for the 
implementation of the proposed decoherence control scheme is
described in detail in \cite{rmp}, in which either 
generation of non-classical states of the radiation \cite{cat,noncl},
and coherent quantum state manipulation \cite{mani}
have been already demonstrated.

A sketch of the set-up is shown in Fig.~1. Its central part is a superconducting
cavity $C$ in a Fabry-Perot configuration, cooled down at about $1$ K.
It sustains two Gaussian field modes with the same spatial structure and orthogonal
linear polarizations. They are separated by a
frequency interval $\Delta = 128 $ kHz around $51.1$ GHz.
The cavity modes can be driven by a tunable classical source $S$. 
These two modes can interact in a controlled way with single, 
velocity-selected, atoms effusing from oven $O$. The atoms
are prepared one at a time in long-lived
(lifetime $\sim 30$ ms), circular Rydberg states, $|e \rangle $ 
(principal quantum number 51) and
$|g \rangle $ (principal quantum number 50), in box $B$. 
The atoms then interact with the cavity, quasi resonant on the 
$e \leftrightarrow g$ transition.
The atom-field coupling is measured by the single-photon Rabi frequency, 
which is time-dependent because of the mode gaussian spatial structure. 
At time $t$, the Rabi frequency writes
$\Omega(t)=\Omega_0\exp\left[-v^2 t^2/w^2\right]$, where 
$w=6$ mm is the mode waist, $\Omega_0/2\pi
=24.5$ kHz \cite{rabi}, and $t=0$ corresponds to the atom 
crossing the cavity axis. The detuning from the cavity mode resonance frequency,
$\delta(t) = \omega_{eg}(t)-\omega$,
can be changed in time in a controlled way using the Stark shift induced
by a uniform electric field applied across the cavity mirrors.
The two-level atom can be manipulated also through microwave pulses generated by
the tunable source $S'$ in a low-$Q$ transverse mode. The final atomic state is 
recorded
by the state-selective detector $D$. 

\begin{figure}[t]
\includegraphics[width=3.5in]{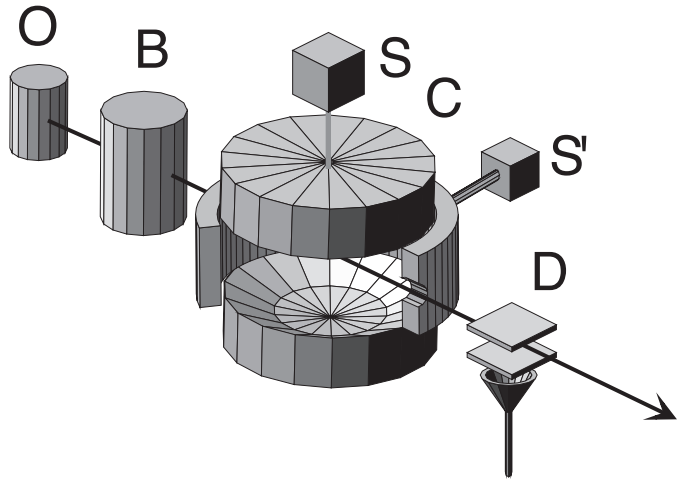}
\caption{ 
Scheme of the experimental apparatus. The Rb atomic 
beam effuses from oven $O$ and circular Rydberg atoms are prepared 
one at a time in box $B$. They cross the high-$Q$ microwave cavity $C$ 
whose state we want to protect, and which can be driven by the source 
$S$. The classical source $S'$ is used for atomic state 
manipulations, and $D$ is the field-ionization detector.
}
\label{fig1}
\end{figure}

The whole protection process is realized by the atom during its transit through
the cavity mode. The first part of the interaction time is used 
for the ``measurement'' of the 
cavity mode parity, while the second part is used for the 
possible state correction. A long interaction time is therefore needed, 
requiring atoms with a moderate
velocity. We are considering here velocities around 
$ v \sim 100$ m/s, which are straightforwardly obtained
in the experiments without need of atomic beam cooling \cite{wign}.
The feedback atoms, sent one by one, are 
initially prepared in the excited state $|e\rangle $. 
The feedback atoms are finally detected in the field ionization
detector D. In a single experimental sequence, we might thus access the
individual quantum trajectory of the cavity field, conditioned to the
atomic detections. We are, however, interested here in an unconditional
decoherence control scheme, able to preserve any quantum state with a
given parity. For this reason, we assume that the information about the
individual atomic state is finally discarded in the data analysis, and
we thus consider only quantum averages of very many individual
trajectories. Note that keeping the atomic state information and
accessing to individual trajectories leads to a different, conditional,
protection scheme which will be discussed elsewhere.
The state of the system just before a generic atom enters the cavity is thus
$|e\rangle \langle e|\rho $, where $\rho$ is the reduced state of the 
cavity mode. We consider only the cavity mode to protect, 
while the other quasi-resonant mode is simply a spectator mode, 
even though it has been taken into account in the numerical simulations 
described in the following Section.

The parity measurement \cite{wign} is performed using a Ramsey interferometry 
scheme \cite{Ramsey}, involving a dispersive interaction in which the 
$e \leftrightarrow g$ transition is light-shifted by the cavity mode \cite{disp}, 
sandwiched between two $\pi/2$ pulses mixing $e$ and $g$ before and after the 
dispersive interaction. The two $\pi/2$ pulses are generated by the 
source $S'$ in a low-$Q$ transverse mode in the cavity structure \cite{rmp}.
In order to minimize a spurious coupling of the Ramsey source $S'$ 
with the superconducting
cavity modes, the atomic transition is shifted far away from the 
cavity resonance by Stark
effect at the time of the $\pi/2$ pulses (see Fig.~2 showing the 
spatial dependence of the atomic detuning within 
the cavity, providing a schematic description of the protection scheme). 
A proper pulse shape
tailoring is used to decrease even further this spurious coupling \cite{wign}.

\begin{figure}[t]
\includegraphics[width=3.5in]{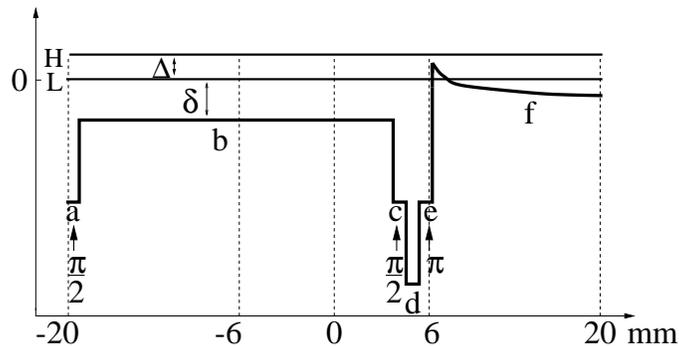}
\caption{
Variation of the atomic detuning $\delta =
\omega_{eg}-\omega$ (thick line) as a 
function of the position within the 
cavity $C$. The two horizontal lines $H$ and $L$ denotes the frequencies
of the two cavity modes, separated by $\Delta = 128$ kHz.
The two vertical dotted lines at $\pm 6$ mm denote the cavity waist.
The variation of the detuning
is obtained by means of the Stark shift 
induced by an electric field applied through the cavity mirrors. The 
various steps of the protection scheme are visible. The two 
$\pi/2$ Ramsey pulses (a) and (c), the 
dispersive $\pi$ phase shift (b), and the phase-tuning Stark-shift 
pulse (d) represent the parity measurement stage of the scheme. 
The $\pi$ pulse $g \rightarrow i$ of step (e)
and the adiabatic transfer step (f) constitutes
instead the correction stage (see text for details).
}
\label{fig2}
\end{figure}

The dispersive interaction between the atom and the 
cavity mode is obtained for sufficiently large atomic detuning, i.e.,
when $|\delta(t)/\Omega(t)| \gg 1$, and for adiabatic 
variations of the parameters. The corresponding Hamiltonian in the 
frame rotating at the cavity mode frequency $\omega$ is then
\begin{equation}
H_{disp}=\hbar \frac{\delta(t)}{2}\left[|e\rangle \langle e| -
|g\rangle \langle g |\right]-
\hbar \frac{\Omega ^{2}(t)}{\delta(t)} \left[|g \rangle \langle g | 
a^{\dagger} a -|e \rangle \langle e | a a^{\dagger}\right]\;.
\label{hdisp}
\end{equation} 
The associated unitary evolution is given by
\begin{equation}
U_{disp}(\phi,\varphi)=|e\rangle 
\langle e | e^{-i\phi/2} e^{-i\varphi a^{\dagger} a }
+ |g\rangle \langle g |  e^{i\phi/2} e^{i\varphi a^{\dagger} a }
\;,
\label{udisp}
\end{equation}
where $\phi = \int dt\left[ \delta(t) + 
\Omega^{2}(t)/\delta(t)\right]$ and $\varphi = \int dt 
\Omega^{2}(t)/\delta(t) $. As shown in 
\cite{protect,jmr,brune96,lutt}, a conditional phase shift per photon 
equal to $\pi$ is needed for a parity measurement, which implies 
adjusting the detuning and the duration of the dispersive interaction 
in such a way that $\varphi = \pi/2$. 

The first stage of the feedback 
loop, describing the parity measurement, is given therefore by the 
transformation
\begin{equation}
|e\rangle \langle e| \rho \rightarrow 
U_{\frac{\pi}{2}}U_{disp}\left(\phi,\frac{\pi}{2}\right)
U_{\frac{\pi}{2}} |e\rangle \langle e| \rho U_{\frac{\pi}{2}}^{\dagger}
U_{disp}\left(\phi,\frac{\pi}{2}\right)^{\dagger}U_{\frac{\pi}{2}}^{\dagger}\;,
\label{trasfo}
\end{equation}
where $ U_{\frac{\pi}{2}} = \left[\left(|e\rangle + |g \rangle 
\right)\langle e|+\left(|g\rangle - |e \rangle 
\right)\langle g|\right]/\sqrt{2}$. Using this fact, the state of the 
system after the step (c) of Fig.~2 can 
be rewritten as
\begin{eqnarray}
&& e^{-i a^{\dagger} a \pi/2 }\left[\left(\frac{1-e^{i\phi}e^{i a^{\dagger} 
a \pi}}{2}\right)|e\rangle + \left(\frac{1+e^{i\phi}e^{i a^{\dagger} 
a \pi}}{2}\right)|g\rangle\right] \rho \left[\langle e|\left(
\frac{1-e^{-i\phi}e^{-i a^{\dagger} 
a \pi}}{2} \right) \right.\nonumber \\
&&+ \left.\langle g|\left(\frac{1+e^{-i\phi}e^{-i a^{\dagger} 
a \pi}}{2} \right)\right] e^{i a^{\dagger} a \pi/2 }.
\label{trasfo2}
\end{eqnarray}
It is evident that the measurement of the cavity mode parity 
is obtained if the phase $\phi$ is appropriately 
adjusted so that $e^{i\phi}= \pm 1$. Each atomic state 
is then unambiguously correlated with a parity eigenvalue. 
The phase $\phi$ can be adjusted to any 
desired value, by strongly detuning the atom from the cavity with a very 
short Stark shift pulse, which has a negligible effect on $\varphi$
(step (d) of Fig.~2).
In the second part of the interaction time, the atom is used to deliver a
photon to the cavity when the ``wrong" parity has been measured. 
The excited state $|e\rangle $ has thus to be 
correlated with the wrong parity component. By choosing $\phi=0$ or $\phi=\pi$,
we can choose which kind of parity eigenstates of the 
cavity mode is protected against photon losses. 

To be specific, we consider from now on the protection of
{\em odd} cavity states. This implies choosing 
$e^{i\phi}= - 1$, so that $|e \rangle $ is associated with even states. 
Finally, note that the overall $\pi/2$ phase space rotation
of the cavity mode in Eq.~(\ref{trasfo2}) can be eliminated by simply 
adjusting the phase of the reference field in $S$, so that the state 
of the system at the end of the 
measurement stage can be written as
\begin{equation}
\left[P_{even}|e\rangle + P_{odd}|g\rangle\right] \rho \left[\langle e|
P_{even} + \langle g| P_{odd}\right] ,
\label{trasfo3}
\end{equation}
where 
\begin{eqnarray}
P_{even}&=& \frac{1+e^{i a^{\dagger} 
a \pi}}{2} \\
P_{odd}&=& \frac{1-e^{i a^{\dagger} 
a \pi}}{2} \label{odd}
\end{eqnarray}
are the projectors onto the even and odd parity eigenspaces, 
respectively.

In the second part of its interaction time with the cavity, 
the atom corrects the 
cavity state component with the wrong parity, by transferring its 
excitation to it. As in \cite{jmr}, this is done using adiabatic 
transfer. When the atomic detuning is adiabatically
changed from a large positive to a large negative value, the system 
remains in the instantaneous dressed state (see \cite{jmr}), realizing 
therefore the transformation
\begin{equation}
|e,n\rangle \rightarrow |g,n+1\rangle \;\;\; \forall n.
\label{adia}
\end{equation}
The photon emission is thus independent of the cavity mode state, an
essential feature for state-independent protection.

If the atom is in state $g$, corresponding to a field measured to be in the
``right" parity state, the opposite adiabatic transfer 
$|g,n\rangle \rightarrow |e,n-1\rangle$ could take place, resulting in a 
spoiled parity. We have thus to get rid of the atom when it exits the parity
measurement in state $g$. This can be 
achieved by tuning the classical source $S'$ on resonance with the 
$g \leftrightarrow i$ transition ($i$ is a lower circular Rydberg state
with principal quantum number $49$ \cite{rmp})
and realizing a $\pi$ pulse $g \rightarrow i$. 
The atom is then ``shelved" in state
$i$, which does not interact with the cavity mode. 
This prevents this unwanted transfer to
occur.

The ideal adiabatic transfer can be formally described by the operator 
\begin{equation}
U_{adia}=|g\rangle \langle e | a^{\dagger}  
\frac{1}{\sqrt{aa^{\dagger}}} + |e\rangle \langle g | a  
\frac{1}{\sqrt{a^{\dagger}a}} +|i\rangle \langle i |\;,
\label{uadia}
\end{equation}
so that the state of the atom-cavity mode system at the end of the 
atomic passage is
\begin{equation}
\left[|g\rangle a^{\dagger}  
\frac{1}{\sqrt{aa^{\dagger}}} P_{even}+ |i\rangle P_{odd}\right] 
\rho \left[
P_{even} \frac{1}{\sqrt{aa^{\dagger}}} a  
\langle g|+  P_{odd} \langle i|\right] .
\label{trasfo4}
\end{equation}
Therefore, in each cycle, the cavity mode is either projected into the 
correct (odd) parity eigenspace, or is corrected via adiabatic 
transfer when it has a wrong (even) parity. The two possibilities could 
be distinguished by detecting the exiting atoms respectively in $i$ 
or in $g$, selecting in this 
way one of the quantum trajectories of the cavity mode 
conditional state. However, a fully quantum feedback has not to rely on the 
classical information provided by the atomic state detection, and it is 
necessarily unconditional. Therefore, we discard
the information about the individual atomic state, and  
tracing over the atom, we get that a generic feedback cycle, i.e., a 
complete atomic passage, can be described by the following map for the 
cavity mode state $\rho$
\begin{equation}
\rho \rightarrow a^{\dagger}  
\frac{1}{\sqrt{aa^{\dagger}}} P_{even} 
\rho P_{even} \frac{1}{\sqrt{aa^{\dagger}}} a  
+  P_{odd} \rho P_{odd} .
\label{trasfo5}
\end{equation}
These are the unitary manipulations characterizing 
the feedback scheme. However, in practice, these manipulations
act simultaneously with the decohering effect of the thermal environment 
described by the master equation (\ref{meq}), and which are responsible 
for the ``errors'' (single photon losses) that the scheme is designed to 
correct for. The resulting evolution is no more unitary and 
described by the simple map of Eq.~(\ref{trasfo5}). The photon 
losses ``contaminate'' the scheme, and the projections onto the parity 
eigenspaces will be no more exact.

\section{Numerical results}

The performance of the proposed protection scheme 
under realistic conditions has been studied by 
solving numerically the master equation describing
the dynamics of the whole system, composed by the 
two non-degenerate high-$Q$ cavity modes
and the two-level atom crossing them. We have 
included also the higher frequency mode (with frequency
$\omega_{H}$ and annihilation operator $a_{H}$)  even though we 
are not interested in its state. 
It is supposed to play only the role
of a spectator in the process. However, in the apparatus described 
in \cite{rmp}, its frequency is very close to the one of the useful mode.
It could be a source of imperfections in the feedback scheme by 
producing uncontrolled phase shifts on the atom during the parity measurement.

We have considered the following master equation for the 
density operator of the whole system $\rho_{T}$
\begin{equation}
\dot{\rho_{T} }= -\frac{i}{\hbar}\left[H,\rho_{T}\right]+
{\cal L}(a)\rho_{T} + {\cal L}(a_{H})\rho_{T},
\label{totmeq}
\end{equation}
where the superoperator ${\cal L}(a)$ has been defined in 
Eq.~(\ref{meq}), and
\begin{eqnarray}
&& H =\hbar \Delta a_{H}^{\dagger} a_{H} +
\hbar \frac{\delta(t)}{2}\left[|e\rangle \langle e| -
|g\rangle \langle g |\right] \label{htot} \\
&&+
i\hbar \Omega(t)
\left[a|e \rangle \langle g | -a^{\dagger} |g \rangle \langle e | \right]+
i\hbar \Omega(t)
\left[a_{H}|e \rangle \langle g | -a_{H}^{\dagger} 
|g \rangle \langle e | \right]\;,
\nonumber 
\end{eqnarray}
is the total Hamiltonian of the system
in the frame rotating at the frequency 
$\omega$ of the mode of interest. We assume that the atoms are sent 
one by one, 
with a spatial separation of $40$ mm. Since the cavity diameter 
is $50$ mm and the mode waist is $6$ mm, this guarantees that the 
atoms interact with the cavity mode one at a time, so that two-atoms
effects are avoided.
Each feedback cycle lasts exactly the time the atom takes to 
cross the cavity region of length $40$ mm around the cavity axis (see 
Fig.~2). Every cycle immediately follows the preceding one, 
starting with the atom entering the interaction region just when the 
preceding one has left it. The feedback atom is always initially 
prepared in state $|e \rangle $  
so that, at the beginning of each cycle, the 
state of the whole system is $|e\rangle \langle e| 
\rho_{out}$, where $\rho_{out}$ is the reduced state of the two modes
at the end of the preceding cycle. 

The master equation has been solved in a truncated Fock 
basis for both modes, using the parameter values  
of the experimental apparatus described in \cite{rmp} (see
the preceding Section). We have also 
assumed that both cavity modes are coupled to a thermal reservoir 
with mean photon number $N=0.8$. This means that thermal excitation 
from the reservoir is not negligible. One might thus expect that
the proposed state protection scheme, designed to correct 
for photon losses only, may not work properly in this case.
We shall see that this is not the case because photon loss is still 
more than twice more probable than thermal excitation. This is 
enough for our feedback scheme to achieve a significant
state protection. 

As it has been already discussed in \cite{protect,jmr}, a 
crucial parameter is the ratio between the time duration of the 
feedback cycle (coinciding with the interaction time of the 
atom) and the relaxation time of the cavity mode $\gamma^{-1}=T_r$. It is 
evident that this ratio has to be as small as possible. Fast atoms
would be preferred. However, the feedback cycle is 
optimal at moderate velocities, since all the 
atomic manipulations have to fit within the 
cavity crossing time. Note that the dispersive step of the parity measurement 
critically depends on the interaction time. In order to fulfill the
$\pi$ phase shift condition ($\varphi = \int dt \Omega^{2}(t)/\delta(t) 
=\pi/2$), the faster the atom, the smaller the detuning 
$\delta$. However, for small values of $\delta$, the dispersive Hamiltonian 
[Eq.~(\ref{hdisp})] is no longer a good approximation
of the total Hamiltonian of Eq.~(\ref{htot}). The correlation between 
the atomic state and the cavity mode parity of Eq.~(\ref{trasfo3}) is 
thus imperfect. We have seen that the best protection results are 
obtained with atomic velocities within the range $80 \div 110$ m/s. 
A clear 
example of quantum state protection in the case of an initial odd 
Schr\"odinger cat state, $|\psi_{1}\rangle\propto |\alpha \rangle -
|-\alpha \rangle $, with $\alpha = 1.8$, is described in Fig.~3, where 
snapshots of the time-evolved Wigner function, both in the presence 
(top) and in the absence (bottom) of protection, are presented.
Note that a proper experimental check of the feedback procedure
would be to map out the cavity state Wigner function, using the 
technique demonstrated in \cite{wign}.

\begin{figure}[t]
\includegraphics[width=3.5in]{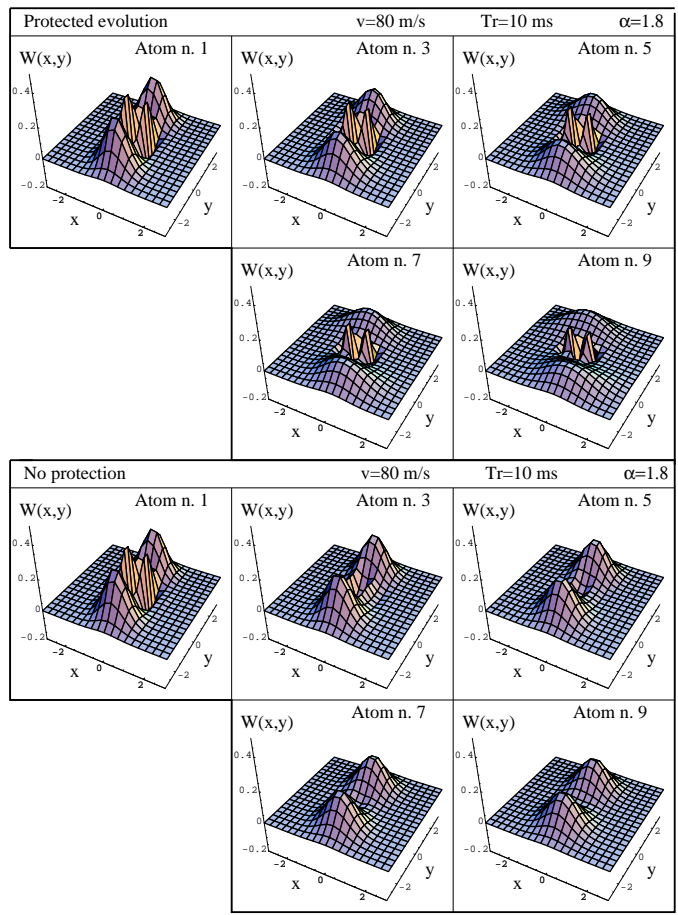}
\caption{
Time evolution of the Wigner function of the cavity mode for an 
initial odd 
Schr\"odinger cat state, $|\psi_{1}\rangle\propto |\alpha \rangle -
|-\alpha \rangle $, with $\alpha = 1.8$, with (top) and without 
(bottom) state protection. The atom velocity is $v=80$ m/s and the 
cavity relaxation time is $T_{r}=10$ ms, while the 
other parameters are given in the text. The elapsed time is measured 
in both cases in terms 
of the number of atomic passages. Time origin is given by the 
exit of the first atom generating the cat 
state out of the interaction region,
and then each atomic passage lasts $500$ $\mu$s.
}
\label{fig3}
\end{figure}

Atom $n.1$ refers to the first atom generating the cat state using the 
scheme already described in \cite{brune96}, and employed in the cat 
state experiment of Ref.~\cite{cat}. This means that, in order to be 
more realistic, we have always considered the protection of the effective 
quantum state generated by the scheme, and not of an initial, ideal, 
quantum state. In the peculiar case of the Schr\"odinger cat state 
of Fig.~3, the generation is obtained assuming the initial coherent 
state $|\alpha \rangle $ injected in the cavity by the source $S$, and 
applying just the parity measurement described in the preceding 
Section. The odd cat state is generated by postselection, when the 
atom is detected in state $|g\rangle $ \cite{brune96}. 
The elapsed time is measured 
(also in the case with no protection, where no atom is used) in terms 
of the number of crossing atoms $n$: the time elapsed from the exit 
of the first atom generating the state out of the interaction region 
is $t_n=L(n-1)/v$, where $L=40$ mm. Fig.~3 refers to an 
atomic velocity $v=80$ m/s and a cavity mode relaxation time 
$T_{r}=10$ ms, which therefore corresponds to $20$ atomic passages.
The comparison with the unprotected evolution indicates a good 
quantum state protection until the ninth atom, even though 
the decoherence time is in this case $t_{dec}=(2\gamma |\alpha 
|^{2})^{-1}=1.54$ ms \cite{milwal2}, corresponding to three atomic 
passages. The 
detuning used in the dispersive stage (step (b) of Fig.~2) is 
$\delta /2\pi =-197$ kHz, while we have used a parabolic variation of the 
detuning $\delta(t)$ around resonance for the adiabatic photon 
transfer (step (f) of Fig.~2) because it turned out to be the 
most effective one. Both the $\pi/2$ Ramsey pulses (steps (a) and (c)), 
and the $\pi$ pulse of step (e) from the source $S'$
had a duration of $1.25$ $\mu$s, 
with their intensity and frequency consistently tuned. 
The Stark shift pulse of step (d), needed to tune the 
phase shift $\phi$ so that $e^{i\phi} = -1$, lasted $1.25$ $\mu$s, 
with a detuning of about $1$ MHz.

A more quantitative description of the 
capabilities of the protection scheme is given by Fig.~4, where the 
time evolution of the fidelity $F(t)=\langle \psi_{1}| \rho(t)
|\psi_{1}\rangle $ and the parity 
$P(t)=\sum_{n}(-1)^{n}\rho_{nn}(t)$ for the odd cat state of Fig.~3 
are shown at fixed atom velocity $v=80$ m/s, and two different 
values of the cavity mode relaxation time, $T_{r}=1,10$ ms.
The stars connected by the dotted line refer to the protected evolution, 
while the diamonds linked by the full line refer to the evolution 
with no protection. The fidelity is appreciably improved 
when $T_{r}=10$ ms, and a small improvement can be seen even 
when $T_{r}=1$ ms. This is not surprising, because a 
single atomic passage lasts $0.5$ ms, which is equal to half 
relaxation time
in this latter case. On the other hand, we can see 
that the proposed protection scheme is actually a very good {\em   
parity preservation scheme}. In fact, the parity of the initially 
generated state (atom n.1) is satisfactorily preserved in time, even in 
the case $T_{r}=1$ ms. In such a case, the initial 
state is far from being an ideal odd cat state
because, due to photon losses, the projection onto 
the odd eigenspace of Eq.~(\ref{odd}) is far from being effectively 
realized. With this respect, our scheme is not fault-tolerant, i.e., 
it does not work perfectly in the presence of losses. Not only the 
projections onto the parity eigenspaces, 
but also the adiabatic transfer in the correction stage is not perfect 
under realistic conditions. Its efficiency when $T_{r}=1$ ms is about 
$90 \% $, and it is due not only to photon losses, but also to the 
fact that the initial positive value of the detuning cannot be taken 
as large as required, because the atom would come close to resonance 
with the high frequency mode and transfer its excitation to it rather 
than to the mode to be protected.

\begin{figure}[t]
\includegraphics[width=3.5in]{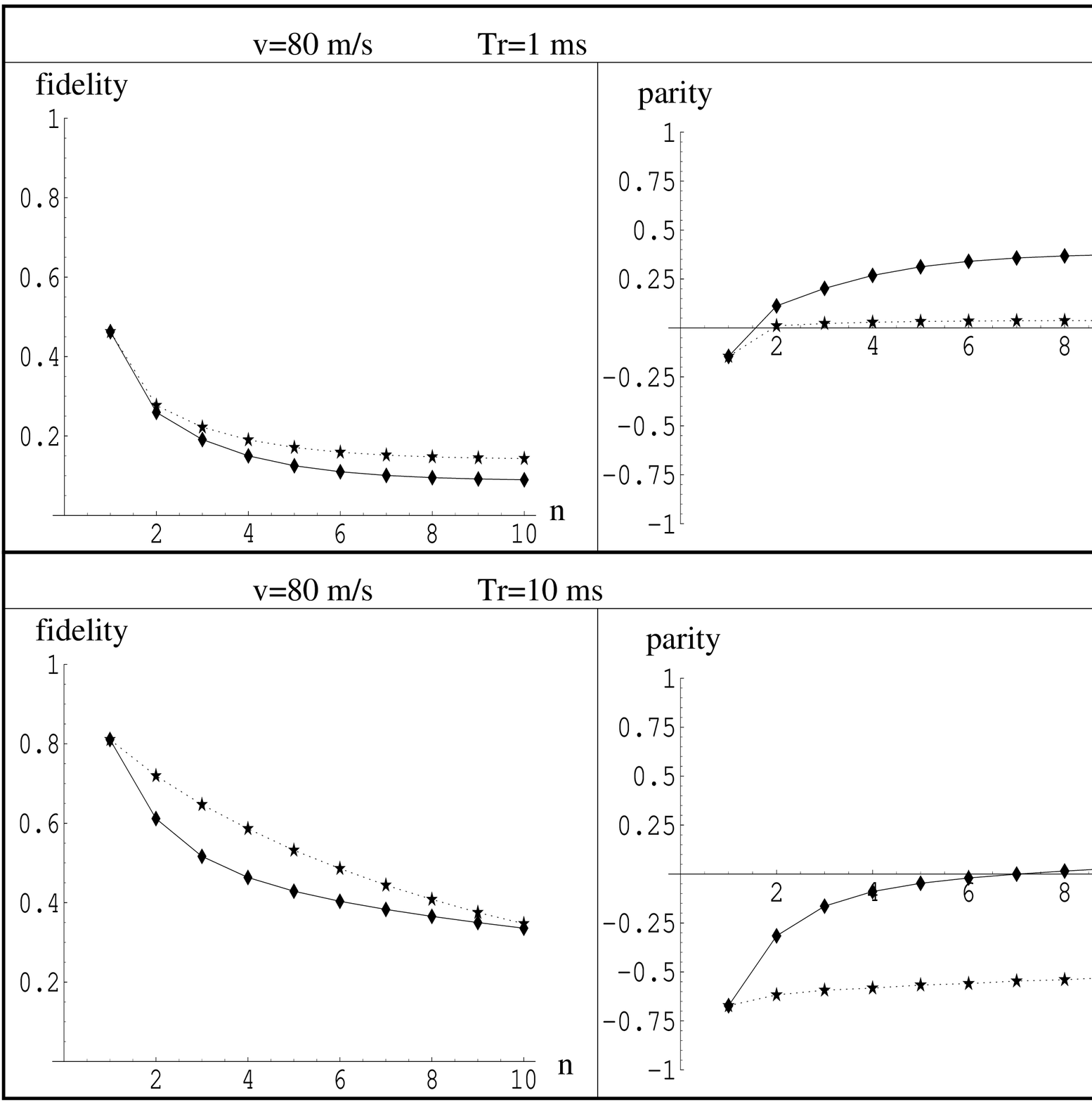}
\caption{
Time evolution of the fidelity and the parity for the
initial odd 
Schr\"odinger cat state of Fig.~3. The stars connected by 
dotted line refer to the protected evolution, 
while the diamonds linked by the full line refer to the evolution 
with no atomic crossing. Parameter values are the same as in Fig.~3 
except that we have considered two values of the relaxation time, 
$T_{r}=1,10$ ms.
}
\label{fig4}
\end{figure}

The most recent experiments with the cavity QED apparatus described 
in \cite{rmp} have been performed with a microwave mode with a 
relaxation time $T_{r}=1$ ms. We have however considered also longer 
relaxation times because there are realistic prospects to achieve
somewhat longer cavity damping times soon. We have also 
investigated the performance 
of the scheme in the case of higher atomic velocities. Protection 
of the odd cat state remains essentially unchanged up 
to $v=110$ m/s, and clearly 
worsens for velocities larger than $150$ m/s. For such velocities the 
dispersive $\pi$ phase shift can be no more realized in a 
satisfactory way.

The scheme is suitable to protect any quantum coherent superposition 
with a given parity, and not only cat states. We have in fact also 
considered the case of an initially generated superposition of two 
Fock states, $| \psi_{2}\rangle =\left(|1\rangle 
+|3\rangle\right)/\sqrt{2}$. The time evolution of the reduced cavity mode 
density matrix in the Fock basis, in the presence of the protection 
scheme, is compared with that with no protection in 
Fig.~5, considering, as in Fig.~3, $T_{r}=10$ ms and
$v=80$ m/s. The other parameter values of the scheme are the same as those
used in the cat state case of Fig.~3.

We have considered also in this case the protection of an initial state 
effectively generated within the apparatus by the first atom.
The generation of the superposition state $| \psi_{2}\rangle =\left(|1\rangle 
+|3\rangle\right)/\sqrt{2}$ can be achieved in the following way. 
Initially the atom is injected in state $|e\rangle $ 
with the cavity mode in the vacuum state (the thermal cavity
field can be erased by sending through it a train of absorbing atoms 
\cite{rmp}). Then, a ``photon pump'' mechanism 
\cite{photpump} can be used 
to transfer photons into the cavity mode. The atomic excitation is 
first transferred to the mode via a resonant atom-cavity interaction. 
The atom is then reset to the excited state (leaving the cavity 
undisturbed) by simultaneously Stark-shifting the atomic levels well out of 
resonance from the cavity mode, and applying a $\pi$ pulse on the transition $e 
\leftrightarrow  g$. By repeating this sequence, one can generate an 
arbitary Fock state $|n\rangle $. In the specific 
case of the state $| \psi_{2}\rangle $, the generation sequence is:
i) resonant $\pi$ pulse yielding $|e,0\rangle \rightarrow 
|g,1\rangle $; ii) Stark-shift and classical $\pi$ pulse giving
$ |g,1\rangle \rightarrow |e,1\rangle $; iii) classical $\pi/2$ pulse on 
the $e\leftrightarrow i $ transition yielding $ |e,1\rangle \rightarrow 
\left(|e,1\rangle +|i,1\rangle\right)/\sqrt{2}$; iv) resonant 
interaction with the cavity, realizing the $\pi$ 
pulse $|e,1\rangle \rightarrow |g,2\rangle $ while leaving the 
$|i\rangle $ component untouched; v) Stark-shift and classical 
$\pi$ pulse giving $ \left(|g,2\rangle + |i,1\rangle\right)/\sqrt{2}
\rightarrow \left(|e,2\rangle + |i,1\rangle\right)/\sqrt{2} $; vi) 
resonant interaction with the cavity mode giving 
$ \left(|e,2\rangle + |i,1\rangle\right)/\sqrt{2}
\rightarrow \left(|g,3\rangle + |i,1\rangle\right)/\sqrt{2} $; vii) 
classical $\pi/2$ pulse on the $g\leftrightarrow i $ 
transition yielding the state 
$\left[|g\rangle \left( |3\rangle - |1\rangle\right)
+ |i\rangle \left(|3\rangle + |1\rangle\right)\right]/2 $, and  
consequent atomic detection in state $|i\rangle $, generating the 
desired superposition $|\psi_{2}\rangle $. Note that in this case, 
the generation can be performed in a very short time and in our 
simulation, we have used a velocity $v=400$ m/s for the 
generating atom.

\begin{figure}[t]
\includegraphics[width=3.5in]{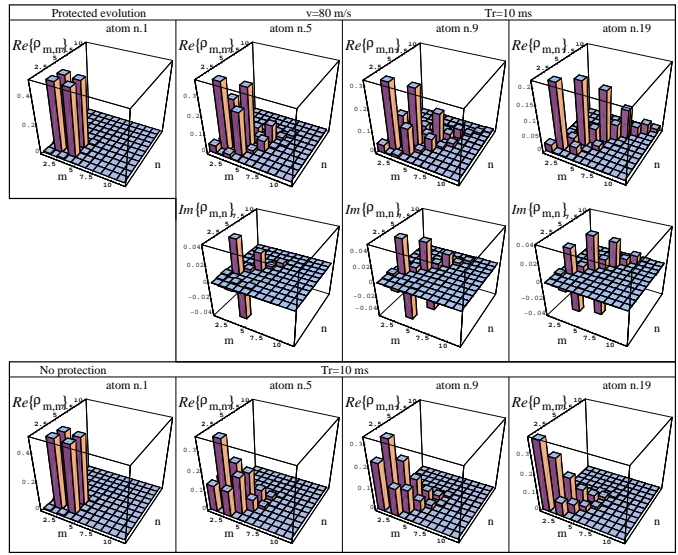}
\caption{
Time evolution of the density matrix in the Fock basis for the 
initial superposition of Fock states $| \psi_{2}\rangle =\left(|1\rangle 
+|3\rangle\right)/\sqrt{2}$ with (top) and without 
(bottom) state protection. The atom velocity is $v=80$ m/s, and the 
cavity relaxation time is $T_{r}=10$ ms, while the 
other parameters are given in the text. The elapsed time is measured 
in both cases in terms 
of the number of atomic passages. 
Time origin is given by the 
exit of the first atom generating the  
state out of the interaction region,
and then each atomic passage lasts $500$ $\mu$s.
}
\label{fig5}
\end{figure}

The density matrix of the protected 
state after $19$ atomic passages
(see Fig.~5) 
clearly shows the effect of 
the feedback-induced ``square-root of phase diffusion'' discussed 
above. This phase diffusion manifests itself at long times, eventually 
driving the cavity mode into a stationary statistical mixture of Fock 
states, corresponding to a rotationally invariant 
Wigner function in phase space \cite{protect} (see also the Wigner function 
of the protected state after $9$ atoms in Fig.~3, where the two Gaussian
peaks start to be stretched by phase diffusion).

We have studied the time evolution 
of the fidelity $F(t) = \langle \psi_{2}|\rho(t) |\psi_{2}\rangle $
and of the parity in this case, both in the presence 
(stars connected with a dotted 
line) and in the absence (diamonds connected by a full line) of 
quantum feedback (see Fig.~6). The comparison between the protected and the
unprotected evolution gives results slightly better
than those of the cat state
(Fig.~4), because the state (especially its parity) is preserved for a
longer time (up to 20 atoms when $T_r = 10$ ms).

\begin{figure}[t]
\includegraphics[width=3.5in]{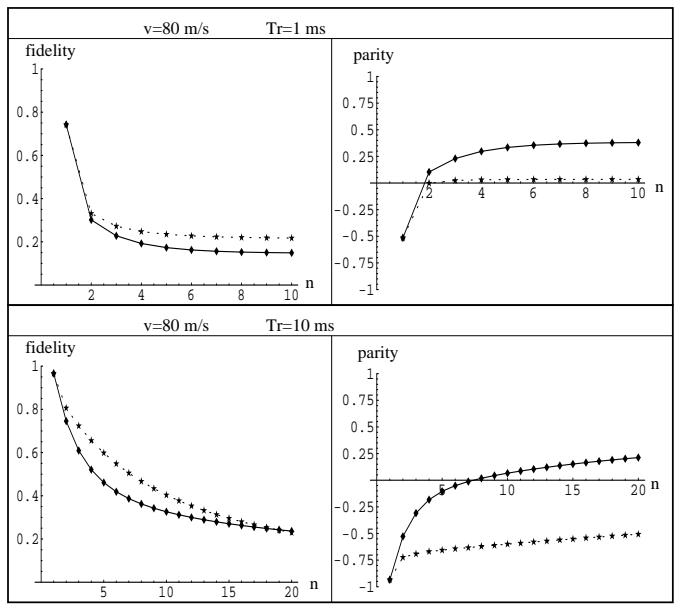}
\caption{
Time evolution of the fidelity and the parity for the
initial superposition of Fock states of Fig.~5. The stars connected by 
the dotted line refer to the protected evolution, 
while the diamonds linked by the full line refer to the evolution 
with no atomic crossing. Parameter values are the same as in Fig.~5, 
except that we have considered two values of the relaxation time, 
$T_{r}=1,10$ ms.
}
\label{fig6}
\end{figure}

As we have seen in Section II, and it is discussed in 
\cite{protect,jmr}, this protection scheme is a stroboscopic 
version of the continuous photodetection feedback scheme described by 
the master equation (\ref{meqfeed}). Therefore, we expect that Fock 
number states of the cavity mode are particularly well protected 
because they are unaffected by the feedback-induced 
phase-diffusion process. We have verified this fact in the 
case of the one photon Fock state and the 
results are shown in Fig.~7, where the time evolution
of the fidelity $F(t)= \rho_{11}(t)$ 
and of the parity $P(t) = \sum_{n}(-1)^{n}\rho_{nn}(t)$ are shown.
The first atom is very fast and generates the $|n=1\rangle $ Fock 
state with a simple resonant interaction $|e,0\rangle \rightarrow 
|g,1\rangle $. Fig.~7 refers
to a cavity relaxation time $T_{r}=1$ ms, 
and the appreciable improvement of the 
fidelity shows that one can demonstrate a significant protection of a 
Fock state using presently available experimental apparata. The 
$|n=1\rangle $ state is easier to protect not only because is not 
affected by phase diffusion, but also because is less sensitive to the 
dispersive step (b) of the scheme.
For this reason, we can use faster atoms and accordingly, 
smaller values of the detuning of step (b). In Fig.~7 we have used 
$v=150$ m/s (with $\delta/2\pi = -109$ kHz) and even $v=200$ m/s 
(with $\delta/2\pi = -73$ kHz).

\begin{figure}[t]
\includegraphics[width=3.5in]{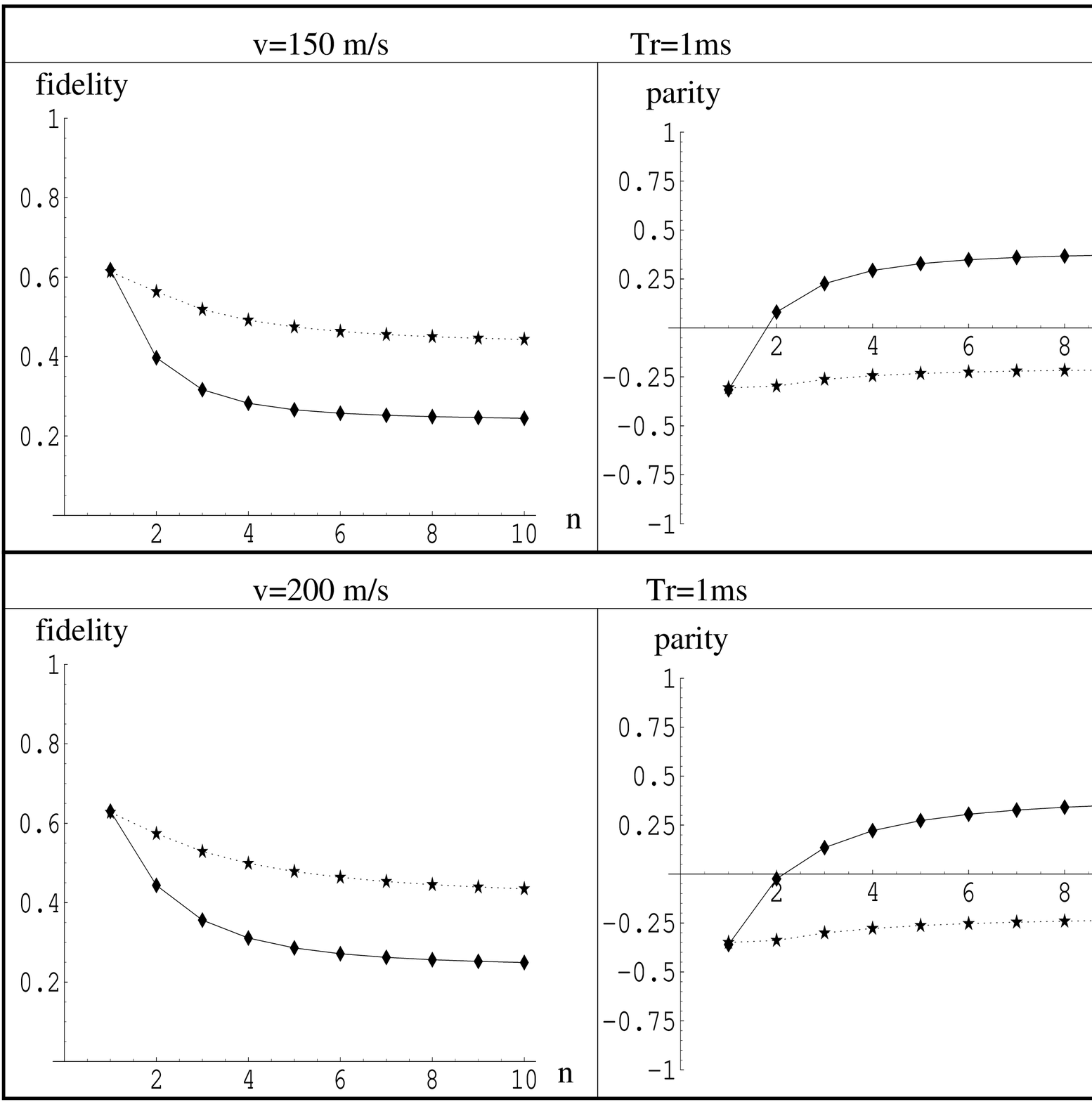}
\caption{
Time evolution of the fidelity and the parity in the case of an 
initial one photon Fock state generated by the first atom. 
The stars connected by the
dotted line refer to the protected evolution, 
while the diamonds linked by the full line refer to the evolution 
with no atomic crossing. The cavity relaxation time is $T_{r}=1$ ms, 
while we have considered two velocities, $v=150$ m/s (top) and 
$v=200$ m/s (bottom). The other parameter values are in the text.
}
\label{fig7}
\end{figure}

Finally, it is also interesting to note that the proposed 
protection scheme could be even used for the {\em generation} of the 
$|n=1\rangle $ Fock state. In fact, as seen in Section II,
state protection is obtained by projecting and eventually restoring 
the component with the desired parity. If one 
starts from a thermal equilibrium state with a low mean photon number 
$N$ so that $\rho_{nn}$ are very small for $n\geq 2$, 
the successive application of the protection scheme for odd states 
will filter out mainly the $|n=1\rangle $ component
which is obviously a stationary state of 
the feedback process \cite{protect}. 

The generation of the one 
photon Fock state starting from a thermal equilibrium state with 
initial cavity mean photon number $N=0.8$ is shown in Fig.~8, 
where, again, the time evolution of the fidelity $F(t)= \rho_{11}(t)$ 
and of the parity are shown (symbols are the same as those of Figs.~4, 
6, and 7). In this case, there is no generation step and the 
first atom is used for feedback already. The plots referring to the case 
with no protection are obviously flat because the 
cavity mode is already at thermal equilibrium. As in Fig.~7, we have 
only considered the case with $T_{r}=1$ ms, and therefore the 
significant difference between the results with and without 
protection in Fig.~8 shows that Fock state filtering from an 
initial thermal distribution could be implemented using the available 
experimental apparatus. Note also that the states involved in the 
protection scheme are rotationally invariant in phase space, 
i.e., are diagonal mixtures in the Fock basis, which are less 
sensitive to the details of the dispersive step (b). Therefore,
we have again used fast feedback atoms and consequently smaller values for the 
detuning of the dispersive step. We have used 
the same values of the single photon state of Fig.~7,
$v=150$ m/s (with $\delta/2\pi = -109$ kHz), and $v=200$ m/s 
(with $\delta/2\pi = -73$ kHz).

\begin{figure}[t]
\includegraphics[width=3.5in]{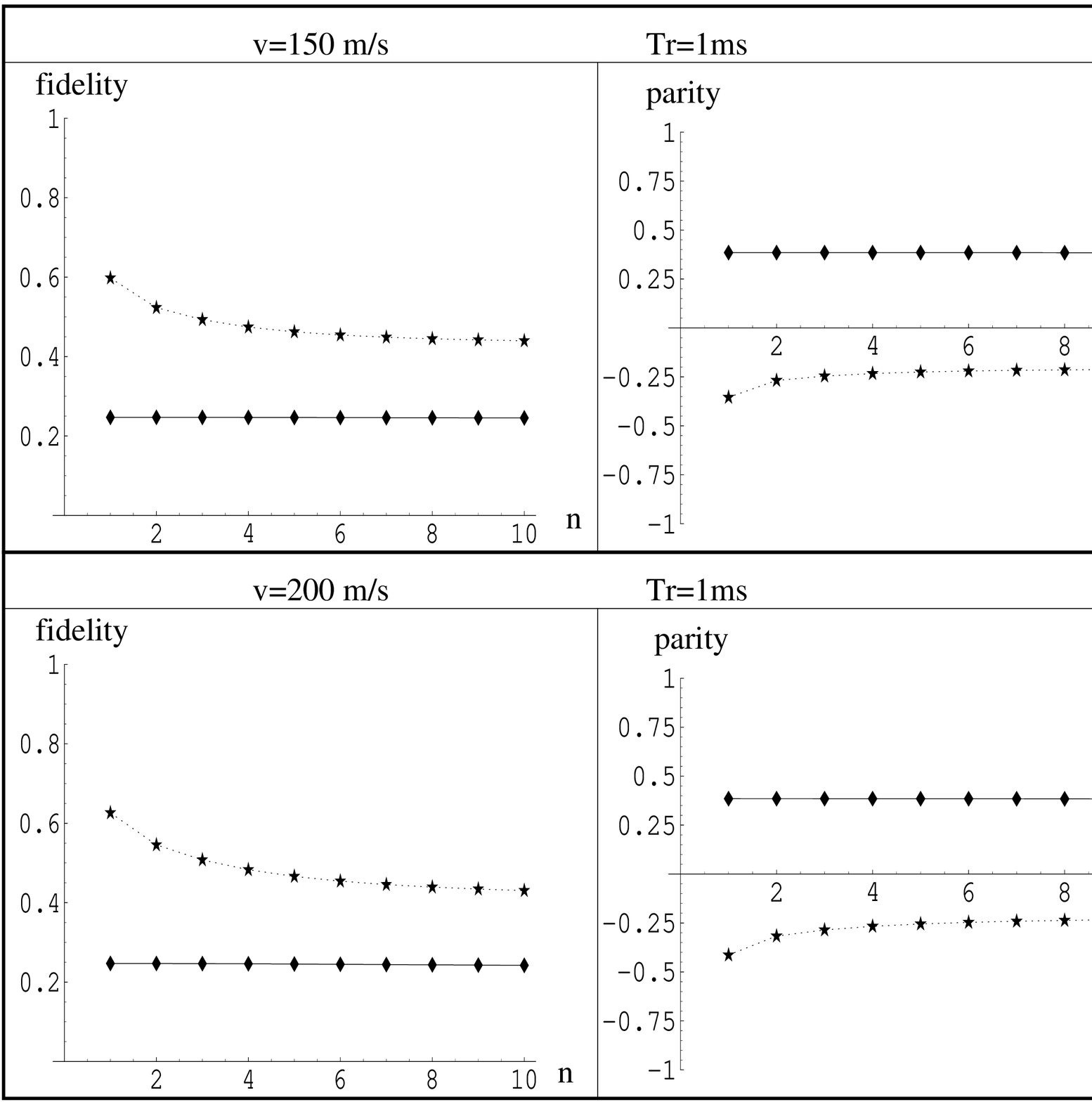}
\caption{ 
Time evolution of the fidelity and the parity in the case when the 
protection scheme is used to ``filter'' the one photon Fock state
from an initial thermal equilibrium state. The stars connected by 
the dotted line refer to the protected evolution, 
while the diamonds linked by the full line refer to the evolution 
with no atomic crossing. The cavity relaxation time is $T_{r}=1$ ms, 
and we have considered two velocities, $v=150$ m/s (top), and 
$v=200$ m/s (bottom). 
The other parameter values are in the text.
}
\label{fig8}
\end{figure}

\section{Conclusions}

We have presented a scheme for the protection of a generic quantum 
state of a cavity mode with a definite parity, against the decohering 
effects of photon losses. The scheme is a further 
improvement of the quantum feedback schemes described in \cite{protect,jmr} and 
is an example of a ``fully quantum feedback'' loop, where sensor, 
controller and actuator are all quantum system \cite{lloyd,lloyd2}. 
In the present scheme, 
all these roles are played by a single atom crossing the cavity mode. 

The scheme presents many analogies with quantum error 
correction codes, even though in our case, there is no explicit 
state encoding. 
In fact, the state to be corrected is already ``encoded'' within a 
parity eigenspace and the error (a single photon loss) maps 
this state into an orthogonal subspace, that is, that with opposite 
parity. Then one corrects the error only within this orthogonal 
subspace. In this respect, the atom plays the role of the 
``error syndrome'' indicating the presence of 
the error. 
Finally the correction is automatically implemented only 
if needed, by means of a kind of C-NOT gate between the atom (control qubit)
and the cavity mode (target qubit represented by the two parity 
eigenspaces).

We have studied the performance of the proposed feedback scheme in 
the case of the cavity QED set up described in \cite{rmp}. We have 
numerically solved the exact master equation by choosing parameter 
values corresponding to those of \cite{rmp}. The 
only simplification adopted is that we have assumed that the circular 
Rydberg atom are prepared one at a time with probability one in the 
set up. In other words, we have assumed a deterministic ``atomic gun''.
This condition is not verified by the present set up, where 
atomic pulses with a mean number of $0.2$ circular atoms per 
pulse are prepared. This fact would lower the efficiency of 
our scheme. 
However, this problem could be circumvented by detecting 
all the feedback atoms exiting the cavity
and postselecting only the events with no 
missing feedback atom (accepting all possible atomic states).

We have considered well separated 
atoms ($40 $ mm) between two successive feedback cycle, just to be 
sure to avoid any two-atom effect. One could increase the efficiency 
of the protection scheme by taking closer feedback atoms. For example, 
if we took 
an atomic separation of $20$ mm, the effective feedback cycle time would be 
halved, and it could still be possible to keep two-atoms effects negligible.

The experimental implementation of the present scheme, especially in 
the simplest cases of a one photon Fock state (Figs.~7 and 8) or a 
superposition of Fock states (Figs.~5 and 6) is feasible with the 
presently available apparatus or with an apparatus with realistic improvements. 
Its demonstration would represent the 
first implementation of decoherence control schemes based on quantum feedback 
ideas, and also the first example of control of a very common, 
and almost unavoidable, source of decoherence such as photon loss.

\section{Acknowledgements} This work has been partially supported by 
the European Union through the IHP program  ``QUEST''.

\end{document}